\documentclass[conference,10pt,table]{IEEEtran}

\usepackage[utf8]{inputenc}
\usepackage{array} 
\usepackage{epsfig}
\usepackage{epstopdf}
\usepackage{times}
\usepackage{framed}
\usepackage{enumitem}
\usepackage{float}
\usepackage{etoolbox}
\usepackage{algorithm}
\usepackage{algorithmicx}
\usepackage{algpseudocode}
\usepackage{amssymb}
\usepackage{amsmath}
\usepackage{multirow}
\usepackage{url}
\usepackage{lipsum}
\usepackage{soul}
\usepackage{caption}
\usepackage{mathtools}
\usepackage{bbm}
\usepackage{booktabs}
\usepackage{xcolor}
\usepackage{makecell}
\usepackage{pifont}
\usepackage{subfigure}
\usepackage[export]{adjustbox} 
\usepackage{latexsym}
\usepackage{cite}
\usepackage[acronym]{glossaries}
\usepackage{multicol}
\usepackage{eurosym}
\usepackage{xspace}
\usepackage{pdfpages}
\usepackage{verbatim}
\usepackage{stfloats}
\usepackage{mathtools}

\newcommand{\vv}[1]{\mathbf{#1}}
\newcommand{\rmpp}{\scriptscriptstyle{\vv{p}}}

\newcommand{\Compl}{\mathbb{C}}
\newcommand{\vvh}[1]{\mathbf{#1^\mathrm{H}}}
\newcommand{\vvt}[1]{\mathbf{#1^\mathrm{T}}}
\newcommand{\herm}{\mathrm{H}}
\newcommand{\tr}{\mathrm{tr}}
\newcommand{\tran}{\mathrm{T}}
\newcommand{\rmA}{\scriptscriptstyle{\mathrm{A}}}
\newcommand{\rmG}{\scriptscriptstyle{\mathrm{G}}}
\newcommand{\rmR}{\scriptscriptstyle{\mathrm{R}}}
\newcommand{\rmD}{\scriptscriptstyle{\mathrm{D}}}

\newcommand{\name}{PAPIR}

\newacronym{ula}{ULA}{uniform linear array}
\newacronym{ap}{AP}{access point}
\newacronym{pdf}{pdf}{probability distribution function}
\newacronym{aod}{AoD}{angle of departure}
\newacronym{aoa}{AoA}{angle of arrival}
\newacronym{ue}{UE}{user equipment}
\newacronym{los}{LoS}{line-of-sight}
\newacronym{pla}{PLA}{planar linear array}
\newacronym[plural=RISs, firstplural=reconfigurable intelligent surfaces (RISs)]{ris}{RIS}{reconfigurable intelligent surface}
\newacronym{sdp}{SDP}{semidefinite programming}
\newacronym{sdr}{SDR}{semidefinite relaxation}
\newacronym{sre}{SRE}{smart radio environment}
\newacronym{snr}{SNR}{signal-to-noise ratio}
\newacronym{toa}{ToA}{time-of-arrival}
\newacronym{doa}{DoA}{direction-of-arrival}
\newacronym{mmse}{MMSE}{minimum mean squared error}
\newacronym{peb}{PEB}{position error bound}
\newacronym{oeb}{OEB}{orientation error bound}
\newacronym{rss}{RSS}{received signal strength}
\newacronym{ml}{ML}{machine learning}
\newacronym{rmse}{RMSE}{root-mean-square error}

\newtheorem{problem}{Problem}

\newcommand{\change}[1]{{\color{black}{#1}}}

\DeclarePairedDelimiter{\norm}{\lVert}{\rVert}

\begin{document}


\setlength{\textfloatsep}{5pt}

\title{PAPIR: Practical RIS-aided Localization \\ via  Statistical User Information}

\author{
    \IEEEauthorblockN{Antonio Albanese\IEEEauthorrefmark{1}\IEEEauthorrefmark{2}, Placido Mursia\IEEEauthorrefmark{1},
    Vincenzo Sciancalepore\IEEEauthorrefmark{1}, Xavier Costa-P\'erez\IEEEauthorrefmark{1}\IEEEauthorrefmark{3}}
    \IEEEauthorblockA{
	\IEEEauthorrefmark{1}NEC Laboratories Europe, Heidelberg, Germany\\
 	\IEEEauthorrefmark{2}Departamento de Ingeniería Telemática, University Carlos III of Madrid, Legan\'es, Spain\\
 	\IEEEauthorrefmark{3}i2cat Foundation and ICREA, Barcelona, Spain
 	\\\{name.surname\}@neclab.eu}
}

\maketitle

\begin{abstract}
The integration of advanced localization techniques in the upcoming next generation networks (B5G/6G) is becoming increasingly important for many use cases comprising contact tracing, natural disasters, terrorist attacks, etc.
Therefore, emerging lightweight and passive technologies that allow accurately controlling the propagation environment, such as \emph{\glspl{ris}}, may help to develop advance positioning solutions relying on channel statistics and  beamforming.
In this paper, we devise \name{}, a practical localization system leveraging on \glspl{ris}
%
by designing a two-stage solution~
building upon prior statistical information on the target \gls{ue} position. \name{} aims at finely estimating the \gls{ue} position by performing statistical beamforming, \gls{doa} and \gls{toa} estimation on a given three-dimensional search space, which is iteratively updated by exploiting the likelihood of the \gls{ue} position.    

\end{abstract}
\begin{IEEEkeywords}
RIS, \gls{doa} estimation, \gls{toa} estimation, Statistical beamforming, \gls{ris}-aided localization. 
\end{IEEEkeywords}

\section{Introduction}
\glsresetall


The radio environment represents the main propagation means that has been deeply analyzed to make wireless communications efficient and reliable. In particular, such a black-box model has been lumped together with advance coding solutions to properly tackle uncontrolled fading issues while still facilitating reasonably-stable channels. Recently, \glspl{ris} appear as the revolutionary and emerging technology bringing the ability of controlling---with passive devices---such propagation environment, via, e.g, backscattering or phase-shifting the incoming electromagnetic waves: this overcomes the traditional adversary perception of the channel thereby turning it into an optimization variable and, in turn, tunable parameter~\cite{DiRenzo2020,Wu2020,Mur20,Cal21,Mur21}.

Manipulating the geometry of the radio propagation has recently drawn vast interest in the scientific community for the design of localization and mapping solutions~\cite{Wymeersch2020,Wymeersch2020VTM,He2020b,Fascista2020}. State-of-the-art \gls{ris}-aided localization employs \glspl{ris} in two alternative ways: \textit{i)} receive mode~\cite{Hu2018}, namely provided with a limited number of RF chains, and \textit{ii)} reflection mode~\cite{Ma2021}. The former techniques aim at localizing a target \gls{ue} fronting the \gls{ris}, e.g., by leveraging on the near-field wavefront curvature~\cite{Guidi2021}. Whereas, the latter category builds upon the reconfigurability of the \glspl{ris} by suitably optimizing the reflection coefficients and allowing a fully passive \gls{ris} design, which is arguably their most attractive operational mode. Indeed, this allows lightening the manufacturing cost and boiling down the overall complexity as well as maximizing deployment flexibility. In this regard, \glspl{ris} can deliver remarkable improvements of one order of magnitude in terms of \gls{peb} with respect to a non-\gls{ris} scenario, given a proper phase design~\cite{elzanaty2020}. Moreover, a \gls{ris} provides better \gls{peb} and \gls{oeb} than a single scatter point~\cite{He2020}.   

Existing \gls{ris}-aided reflection mode localization algorithms leverage on direct maximum-likelihood estimate of the 
\gls{ue} position, which is known to be computationally expensive~\cite{elzanaty2020}, or fingerprinting solutions based on \gls{rss} measurements at the \gls{ue}-side. Indeed, \glspl{ris} can exacerbate the \gls{rss} differences among different locations, thus improving the localization accuracy~\cite{Zhang2021}. In this regard, the performance of fingerprinting can be enhanced by the aid of \gls{ml} techniques via feature selection, which prunes the large state space of the \gls{ris} and reduces the overall complexity of the position estimation~\cite{nguyen2020}. However, the above-mentioned techniques lack of practical assumptions on the unknown \gls{ue} location. Since proper phase shift configuration of \glspl{ris} operating in reflection mode requires the \gls{ue} position, which is exactly the
quantity to be estimated, a chicken-egg problem raises.

In this paper, we propose a novel two-stage \gls{ris}-aided localization algorithm denoted as \name{}, namely PAssive PosItioning with \gls{ris}, which retrieves the \gls{toa} and the \gls{doa} of reference signals sent by the \gls{ue} to estimate its position. We assume the \gls{ris} to be in reflection mode and we optimize its reflection coefficients based on some (possibly) coarse prior information on the \gls{ue} position. To the best of our knowledge, this is the first work presenting a practical and efficient solution able to localize a \gls{ue} with realistic assumptions on its position. We present a thorough numerical evaluation to assess the performance of the proposed scheme in terms of localization accuracy.

\noindent\textbf{Notation.} We use $\Compl^{n}$ and $\Compl^{m\times n}$ to represent the sets of $n$-dimensional complex vectors and $m\times n$ complex matrices, respectively. Vectors are denoted by default as column vectors. We let $(\cdot)^{\vvh{}}$, $(\cdot)^{\vvt{}}$, and $(\cdot)^*$ to denote the Hermitian, transpose, and conjugate operators, respectively. $\norm{\cdot}$ is the $\mathrm{L}2$-norm of a vector, whereas $\vv{I}_N$ and $\otimes$ are the $N$-dimensional identity matrix and the Kronecker product, respectively.

\section{System Model}

\begin{figure}[t!]
    \centering
    \vspace{5.3mm}
    \includegraphics[clip,width=\linewidth]{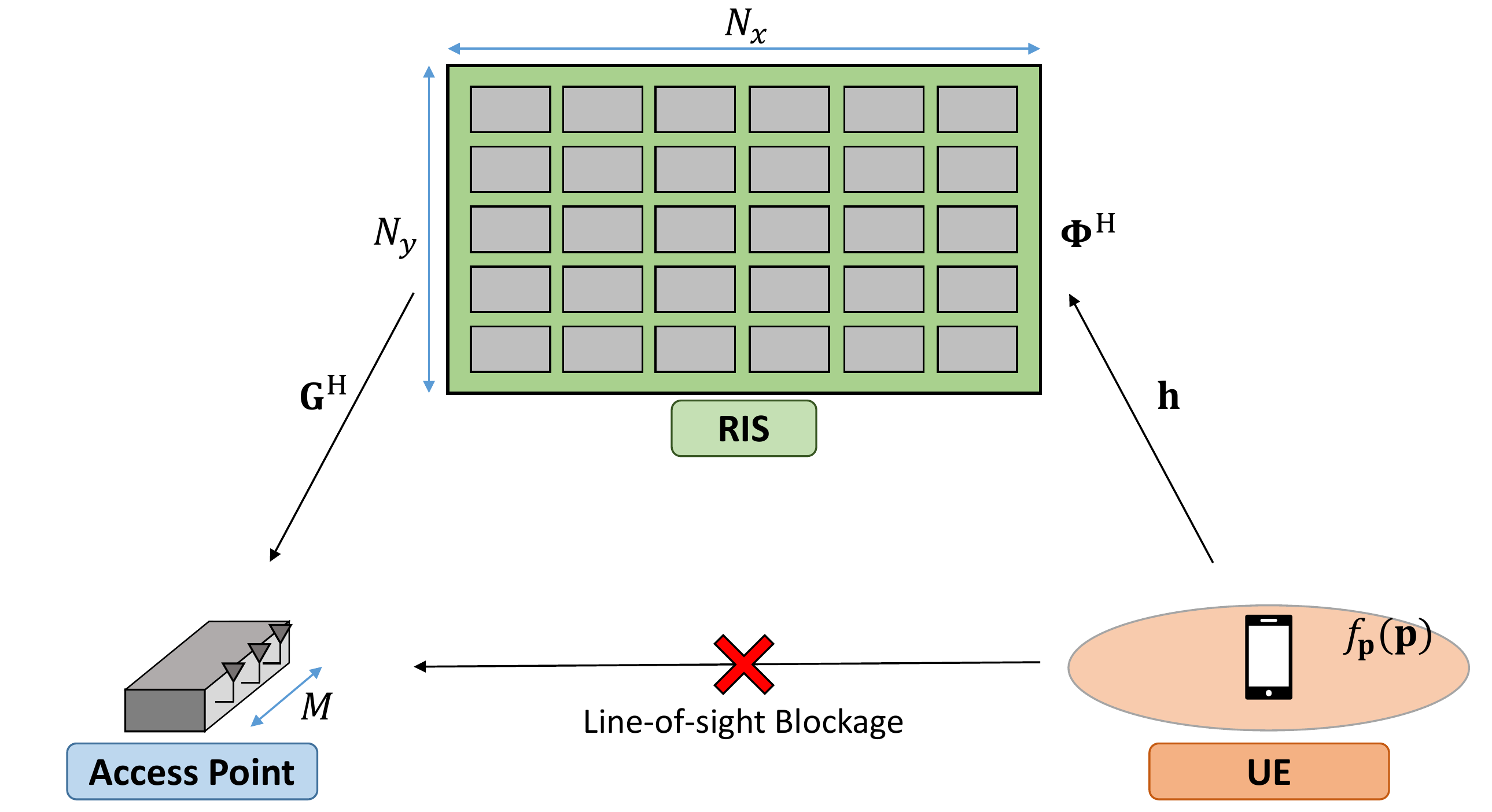}
    \caption{Localization scenario}
    \label{fig:scenario}
\end{figure}

We consider the scenario depicted in Fig.~\ref{fig:scenario} where an \gls{ap} equipped with $M$ antennas performs the localization of a target single-antenna \gls{ue} with the aid of a \gls{ris} consisting of $N$ elements. Specifically, the \gls{ue} transmits uplink reference signals to the \gls{ap} who then processes the received signal and estimates the \gls{ue} position. We model the \gls{ap} as a \gls{ula}, while the \gls{ris} is assumed to be a \gls{pla} with $N_x$ and $N_y$ elements along the $x$ and $y$ axis, respectively, with $N=N_xN_y$. The \gls{ap} is located at the origin of our reference system and the \gls{ris} array center has coordinates $\vv{p}_{\rmR} \in \mathbb{R}^3$, while only a given \gls{pdf} of the position of the \gls{ue} $\vv{p} \in \mathbb{R}^3$ is known and given by $f_{\rmpp}(\vv{p})$. We assume that there is no direct link between the \gls{ue} and the \gls{ap}, so that communication between the two must be established via the path reflected upon the \gls{ris}. The latter is decomposed into the \gls{los} channel $\vv{h}(\vv{p}) \in \mathbb{C}^{N  }$ through which the \gls{ris} reflects the impinging signal towards the \gls{ap} for a given \gls{ue} position $\vv{p}$, and the \gls{ap}-\gls{ris} \gls{los} link denoted by $\vv{G} \in \mathbb{C}^{N \times M}$. We assume that all channels follow a quasi-static flat-fading model and thus remain constant over the transmission time of the reference signals. \change{Moreover, we employ millimeter wave frequencies, which are characterized by sparse multipath even in complex propagation conditions such as dense urban or indoor environments.} Let $\psi_{\rmD}$ and $\psi_{\rmA}$ indicate the \gls{aod} from the \gls{ris} and the \gls{aoa} at the \gls{ap}, respectively. Hence, the \gls{los} \gls{ris}-\gls{ap} channel $\vv{G}$ is defined as 
\begin{equation}
    \vv{G} \triangleq \sqrt{\gamma_{\rmG}} \, \vv{b}(\psi_{\rmD}) \vvh{a(\psi_{\rmA})},
    \label{eq:G}
\end{equation}
where $\gamma_{\rmG} \triangleq d_{\rmG}^{-\beta}$ is the channel power gain with $d_{\rmG}$ the distance between the \gls{ris} and the \gls{ap} and $\beta$ the pathloss exponent, $\vv{b}(\psi_{\rmD}) \in \mathbb{C}^{N  }$ is the \gls{pla} response vector at the \gls{ris} for the steering angle $\psi_{\rmD}$, and $\vv{a}(\psi_A)\in \mathbb{C}^{M  }$ is the \gls{ula} response vector at the \gls{ap} for the steering angle $\psi_A$. The former is given by
\begin{align}
    \vv{b}(\psi_{\rmD}) \triangleq \, &  \vv{b}_z (\psi_{\rmD,z}, \psi_{\rmD,x}) \otimes \vv{b}_x (\psi_{\rmD,z}, \psi_{\rmD,x}) \label{eq:pla}\\
    = \,&[1, e^{j2\pi\delta\sin(\psi_{\rmD,z})\cos(\psi_{\rmD,x})}, \dots, \nonumber\\
    &e^{j2\pi\delta(N_y -1)\sin(\psi_{\rmD,z})\cos(\psi_{\rmD,x})}]^\mathrm{T} \nonumber \\
    &\otimes[1, e^{j2\pi\delta\sin(\psi_{\rmD,x})\cos(\psi_{\rmD,z})}, \dots, \nonumber \\
    &e^{j2\pi\delta(N_x -1)\sin(\psi_{\rmD,x})\cos(\psi_{\rmD,z})}]^\mathrm{T},
\end{align}
where $\psi_{\rmD,z}$ and $\psi_{\rmD,x}$ are the azimuth and elevation \gls{aod}, respectively, and \change{$\delta=0.5$} is the antenna spacing-wavelength ratio. In a similar way, the \gls{ula} response at the \gls{ap} is defined as s
\begin{equation}
    \vv{a}(\psi_A) \triangleq [1, e^{j2\pi\delta\cos(\psi_A)}, \dots, e^{j2\pi\delta(M-1)\cos(\psi_A)}]^\mathrm{T}.
\end{equation}
Note that the coordinates of the \gls{ris} position $\vv{p}_{\rmR} = [p_{\rmR,x},p_{\rmR,y},p_{\rmR,z}]^{\tran}$ are expressed as $p_{\rmR,x} = \norm{\vv{p_{\rmR}}}\cos(\psi_{\rmD,z})\cos(\psi_{\rmD,x})$, $p_{\rmR,y} = \norm{\vv{p_{\rmR}}}\cos(\psi_{\rmD,z})\sin(\psi_{\rmD,x})$, and $p_{\rmR,z} = \norm{\vv{p_{\rmR}}}\sin(\psi_{\rmD,z})$, respectively. \change{The \gls{ue}-\gls{ris} channel for a given (unknown) \gls{ue} position $\vv{p}$ reads as}
\begin{equation}\label{eq:h_ue}
    \vv{h}(\vv{p}) \triangleq \sqrt{\gamma} \, \vv{b}(\theta),
\end{equation}
where $\gamma \triangleq d^{-\beta}$ is the channel power gain with $d = \norm{\vv{p}_{\rmR} - \vv{p}}$ the Eucledian distance between the \gls{ue} and the \gls{ris}, and $\vv{b}(\theta)$ is the \gls{pla} response vector of the \gls{ris} for the steering angle $\theta$, i.e., the \gls{aoa} of the \gls{los} \gls{ue}-\gls{ris} path, as in Eq.~\eqref{eq:pla}. Note that both $d$ and $\theta$ are random variables, which are obtained via proper transformation of the \gls{ue} position $\vv{p}$, and are thus characterized by the \gls{pdf} of the \gls{ue} distribution $f_{\rmpp}(\vv{p})$.

The received uplink signal at the \gls{ap} in time slot $n$ is thus defined as
\begin{equation}\label{eq:y}
    \vv{y}(\vv{p},n) \triangleq \sqrt{P}\, \vvh{G} \vv{\Phi}(n)^{\herm} \vv{h}(\vv{p}) \,  s(n) + \vv{n}(n)\in \Compl^{M },
\end{equation}
where $P$ is the transmit power at the \gls{ue}, $\vv{\Phi}(n) = \mathrm{diag}[\alpha_1(n) e^{j\phi_1(n)}, \dots, \alpha_N(n) e^{j\phi_N(n)}]$ with $\phi_i(n) \in [0, 2\pi]$ and $|\alpha_i(n)|^2 \leq 1$, $\forall i$ indicates the phase shifts and amplitude attenuation introduced by the \gls{ris} in time slot $n$ (to be optimized), while $s(n)\in \Compl$ is the (known) transmit signal with $|s(n)|^2=1$, $\forall n$, and $\vv{n}(n)\in \Compl^{M }$ is the additive white Gaussian noise term distributed as $\mathcal{CN}(0,\sigma^2\vv{I}_M)$, $\forall n$. For the sake of simplicity, we assume that the reflections at the \gls{ris} are ideal, i.e., that the amplitude attenuations $\{\alpha_i(n)\}_{i=1}^N$ and the phase shifts $\{\phi_i(n)\}_{i=1}^N$ can be independently optimized\footnote{\change{In real prototypes, amplitude attenuations and phase shifts are dependent and affect the RIS beamforming performance~\cite{Abeywickrama2020}. However, this issue is out of the scope of this work.}}. Lastly, we define the received sum \gls{snr} at the \gls{ap} antennas in time slot $n$ as
\begin{equation}\label{eq:snr}
    \mathrm{SNR}(\vv{p},n) \triangleq P\, \frac{\norm{\vvh{G} \vv{\Phi}(n)^{\herm} \vv{h}(\vv{p})}^2}{\sigma^2},
\end{equation}

\change{where $\vv{G}$ is fully defined by knowing the coordinates of the AP and the RIS as per Eq.~\eqref{eq:G}, $\vv{\Phi}$ is set during the RIS optimization phase and $\vv{h}$ is a byproduct of the localization procedure.}

\section{RIS-Aided Localization}

Given the model for the channel between the \gls{ue} and the \gls{ris} in Eq.~\eqref{eq:h_ue}, the problem of determining the \gls{ue} location reduces to the estimation of the steering angle $\theta$ and the distance from the \gls{ris} $d$. In this paper, we propose to suitably configure the \gls{ris} to reflect the incoming signal from the \gls{ue} to the \gls{ap}, who then performs \gls{doa} estimation to determine $\theta$, and \gls{toa} estimation to obtain $d$. In this respect, we optimize the \gls{ris} configuration $\vv{\Phi}(n)$ at each time slot $n$ in order to maximize the \gls{snr} at the \gls{ap} antennas in Eq.~\eqref{eq:snr} for any given position of the target \gls{ue} $\vv{p}$. Notably, $\vv{p}$ is exactly the quantity we aim at estimating to perform the localization of the \gls{ue}, thus giving rise to a chicken-egg problem. Therefore, we exploit the prior statistical information about the \gls{ue} position, which is given by $f_{\rmpp}(\vv{p})$, and design an efficient iterative algorithm, denoted as \name{}, aiming at obtaining a \gls{ris} configuration that is robust against the uncertainty in the \gls{ue} position. Specifically, we adopt the minimum received SNR among the possible \gls{ue} positions as objective function.

\subsection{DoA Estimation}

The task of optimizing the \gls{ris} configuration for any given pdf of the \gls{ue} position $f_{\rmpp}(\vv{p})$ is particularly challenging. Indeed, $f_{\rmpp}(\vv{p})$ provides only coarse information on the desired direction of reflection at the \gls{ris}, thus leading to low received SNR and ultimately poor estimation accuracy. To circumvent this problem, we propose to divide the total three-dimensional area of possible \gls{ue} positions $\mathcal{A}^{(n)}$ identified by $f_{\rmpp}(\vv{p})$ into a set of $N_A$ subareas $\{\mathcal{A}^{(n)}_1,\ldots,\mathcal{A}^{(n)}_{N_A}\}$ of equal size $|\mathcal{A}^{(n)}|/N_A$ with a given degree of overlap $\Delta$ between two adjacent areas such that
\begin{equation}
    \mathcal{A}^{(n)}=\bigcup_{\ell=1}^{N_A} \mathcal{A}^{(n)}_{\ell}.
\end{equation}
The \gls{ap} probes each of the subareas by suitably configuring the \gls{ris}, thus producing $N_A$ \gls{doa} estimates $\{\hat{\theta}_1,\ldots,\hat{\theta}_{N_A}\}$. 
The \gls{ap} then defines the next probing space $\mathcal{A}^{(n+1)}$ as the area defined by the union of the subareas corresponding to the $N_A-1$ \gls{doa} estimates that maximize the likelihood of the \gls{ue} position, which is defined by the pdf $f_{\rmpp}(\vv{p})$.\footnote{Without loss of generality, we assume that $N_A\geq 3$.} This procedure is repeated until the difference among the \gls{doa} estimates of two consecutive iterations of \name{} is less than a given threshold $\epsilon$. Note that both $\epsilon$ and $N_A$ regulate a trade-off between accuracy and estimation time and thus need to be properly designed. Let $f_{\rmpp}^{(n)}(\vv{p})$ be the pdf of \gls{ue} positions corresponding to $\mathcal{A}^{(n)}_{\ell}$, with $\ell = n\;\mathrm{mod}\; N_A$. Hence, at each time slot $n$, the \gls{ris} is configured to maximize the minimum SNR among the possible \gls{ue} positions in the area identified by $f_{\rmpp}^{(n)}(\vv{p})$, which decreases in size between consecutive probing spaces. This allows us to perform increasingly more selective beamforming, which will give rise to a very sharp increase in SNR when the \gls{ris} is pointing in the direction of the \gls{ue}, compared to when it is pointing in the incorrect direction.

Let us focus on a specific time slot $n$: here, to ease the notation we drop the time index $n$ and we define $\vv{v} \triangleq [\alpha_1 e^{-j\phi_1}, \dots, \alpha_N e^{-j\phi_N}]^{\tran}$, with $\vv{\Phi} = \mathrm{diag}(\vv{v}^{\herm})$, and the equivalent uplink channel $\vv{H}(\vv{p}) \triangleq \mathrm{diag}(\vvh{h(p)})\vv{G}$ such that we can reformulate the \gls{snr} in Eq.~\eqref{eq:snr} as
\begin{equation}
    \mathrm{SNR}(\vv{p}) = P\, \frac{\norm{\vv{v}^{\herm} \vv{H}(\vv{p}) }^2}{\sigma^2}.
\end{equation}
Hence, we formulate the following optimization problem
\begin{problem}[Statistical \gls{ris} beamforming]\label{problem:max_snr}
\begin{align}
   \displaystyle \max_{\vv{v}} & \displaystyle \min_{\vv{p} \sim f_{\rmpp}(\vv{p})} P \ \frac{\norm{\vv{v}^{\herm} \vv{H}(\vv{p}) }^2}{\sigma^2} \\
   \displaystyle \textup{s.t.} & \ \ \displaystyle |v_i|^2 \leq  1 \quad \forall i, \label{eq:phi_con} 
\end{align}
\end{problem}
whose solution provides the \gls{ris} configuration maximizing the minimum receive SNR over the possible \gls{ue} positions $\vv{p}$, which are determined by $f_{\rmpp}(\vv{p})$. However, note that Problem~\ref{problem:max_snr} is highly complex to tackle due to the non-convex maximization of a quadratic function in $\vv{v}$ and the general expression of the \gls{ue} distribution $f_{\rmpp}(\vv{p})$. In this regard, we propose to employ Monte Carlo sampling and \gls{sdr} in order to find a simple yet effective solution to the above problem.

Let us draw $T$ sample points $\{\vv{p}_t\}_{t=1}^T$ from $f_{\rmpp}(\vv{p})$ and let $\vv{V}\triangleq \vv{v}\vvh{v}$, and $\overline{\vv{H}}(\vv{p})\triangleq\vv{H}(\vv{p})\vvh{H(p)}$, such that Problem~\ref{problem:max_snr} is reformulated as
\begin{problem}[Statistical \gls{ris} beamforming with SDR]\label{problem:max_snr_SDR}
\begin{align}
   \max_{\vv{V}\succeq\vv{0}} & \min_{\{\vv{p}_t\}_{t=1}^T} \tr(\vv{\overline{H}}(\vv{p}_t)\vv{V}) \\
   \textup{s.t.} & \ \ \mathrm{diag}(\vv{V}) \leq 1 \\
   & \ \ \mathrm{rank}(\vv{V})=1\label{eq:rank_constraint},
\end{align}
\end{problem}
where we omit the scaling factor $P/\sigma^2_n$ as it is irrelevant for the optimization. Problem~\ref{problem:max_snr_SDR} can be solved in its relaxed form, i.e., when ignoring the non-convex rank constraint in Eq.~\eqref{eq:rank_constraint}, by employing canonical semidefinite programming such as CVX. Let $\vv{V}^{\star}$ denote the optimal solution of the relaxed version of Problem~\ref{problem:max_snr_SDR}, then we can recover a suboptimal solution to Problem~\ref{problem:max_snr}, namely $\vv{v}^{\star}$, by Gaussian randomization.

Given the \gls{ris} configuration, i.e., $\vv{\Phi}=\mathrm{diag}(\vvh{(v^{\star})})$, the \gls{doa} of the \gls{ue} $\theta$ is estimated via a suitable MUSIC-based processing at the \gls{ap} which is detailed as follows. However, the MUSIC procedure cannot be directly applied to the received signal $\vv{y}$ in Eq.~\eqref{eq:y} since it would estimate the \gls{doa} of the equivalent uplink channel $\vvh{G}\vvh{\Phi}\vv{h}$, which includes the target \gls{doa} $\theta$ plus the effect of the phase shift applied at the \gls{ris} and the channel path towards the \gls{ap} $\vv{G}$. Moreover, the receive signal $\vv{y}$ is a vector of dimension $M$ whereas the estimation target $\vv{h}(\vv{p})$ is a vector of dimension $N$.\footnote{Note that we assume $M<N$, i.e., that the number of \gls{ap} antennas is less than the number of \gls{ris} elements.} To this end, we firstly equalize the received signal $\vv{y}$ by employing \gls{mmse} filtering as
\begin{align}
    \vv{x}(\vv{p}) & = \frac{1}{\sqrt{P}}\ \vv{W} \vv{y}(\vv{p})s^* \label{eq:x}\\
    & = \vv{W}\vvh{G}\vvh{\Phi}\vv{h}(\vv{p}) + \overline{\vv{n}} \in \Compl^{N }
\end{align}
where we set
\begin{align}\label{eq:W}
    \vv{W} \triangleq \left(\vv{\Phi}\vv{G}\vvh{G}\vvh{\Phi}+\frac{\sigma_n^2}{P}\vv{I}_N\right)^{-1}\vv{\Phi}\vv{G} \in \Compl^{N\times M},
\end{align}
with the equivalent noise term $\overline{\vv{n}} \triangleq \frac{1}{\sqrt{P}} \ \vv{W}\vv{n} s^*$. Note that such filtering procedure results in projecting the receive signal onto the subspace spanned by the \gls{ue} channel $\vv{h}(\vv{p})$. Moreover, it does not alter the rank of said subspace, which remains equal to $1$. Hence, the filtered signal $\vv{x}(\vv{p})$ can be fed to the MUSIC procedure, which estimates the direction of arrival $\hat{\theta}$ by exploiting the orthogonality between the subspace corresponding to $\vv{h}(\vv{p})$ and the noise subspace. 

\subsection{ToA Estimation}

The \gls{toa} is calculated
by exploiting the ideal autocorrelation property of the transmit sequences. To this end, we set the sequence $\{s(n)\}_{n}$ to a Constant-Amplitude Zero-Autocorrelation (CAZAC) sequence such as the Zadoff-Chu
(ZC). Without loss of generality, we assume that the total transmission time $L$ is an odd number such that we set
\begin{align}
    s(n) = e^{-j\pi \frac{n(n+1)}{L}}, \quad n=1,\ldots,L.
\end{align}
Hence, we have that the autocorrelation of the transmit signal at lag $m$ is given by
\begin{align}
    R_{ss}(m)& \triangleq \sum_{n=1}^Ls(n)s^*(n+m) = D(m)
\end{align}
where $D(m)$ is the Dirac delta function. The \gls{toa} of the transmit sequence is thus estimated as $\tau = m^{\star} / \Delta f$, where $m^{\star}$ is the lag corresponding to the peak in the cross-correlation matrix of the received signal with the transmit sequence $\vv{R}_{\vv{y}s}$ and $\Delta f$ is the sampling frequency.\footnote{In general, the estimation precision of the \gls{toa} can be improved by upsampling both the received signal and the transmit sequence before computing the cross-correlation matrix between the two.} Note that since the reflection upon the \gls{ris} is instantaneous and does not increase the total \gls{toa}, the distance from the \gls{ue} to the \gls{ris} $d$ is estimated as $\hat{d} = \tau c - d_{\rmG}$, where $c$ is the speed of light. Finally, the proposed \name{} procedure is formalized in Algorithm~\ref{alg:A1}. \change{Its complexity is dictated by the solution of the relaxed version of Problem~\ref{problem:max_snr_SDR}, i.e., standard semidefinite programming, whose convergence is guaranteed thanks to the convex nature of the problem at hand. Overall, the proposed method is observed to converge in less than $10$ iterations each one having a complexity of $\mathcal{O}(\sqrt{N}(N^6+N^3))$.} 

\begin{algorithm}[t!]
  \caption{\name: Passive Positioning with \gls{ris}}\label{alg:A1}
  \begin{algorithmic}[1]
     \State Data: $f_{\rmpp}(\vv{p})$, $\mathcal{A}$, $N_A$, $\Delta$, $\Delta f$, $d_{\rmG}$, $\epsilon$ and $T$
     \State Set $\mathcal{A}^{(1)} = \mathcal{A}$, $\epsilon>0$, and $n=\ell=1$
     \State Initialize $\hat{\theta}^{(0)} \neq \hat{\theta}^{(1)}$
     \State Divide  $\mathcal{A}^{(1)}$ into $N_A$ equal subareas $\{\mathcal{A}^{(1)}_1,\ldots,\mathcal{A}^{(1)}_{N_A}\}$ with $\Delta$ overlap
     \While { $|\hat{\theta}^{(n-1)}-\hat{\theta}^{(n)}|>\epsilon$ }
     \For {$\ell=1,\ldots,N_A$}
     \State Calculate $f_{\rmpp}^{(n)}(\vv{p})$ over the subarea $\mathcal{A}^{(n)}_{\ell}$
     \State Draw $T$ points $\{\vv{p}_t\}_{t=1}^T$ according to $f_{\rmpp}^{(n)}(\vv{p})$
     \State Obtain $\vv{V}^{\star}$ by solving the relaxed version of  Problem~\ref{problem:max_snr_SDR}
     \State Obtain $\vv{v}^{\star}$ from $\vv{V}^{\star}$ via Gaussian randomization
     \State Set $\vv{\Phi} = \mathrm{diag}((\vv{v}^{\star})^{\herm})$ and collect $\vv{y}$
     \State Set $\vv{W}$ as in Eq.~\eqref{eq:W} and $\vv{x}$ as in Eq.~\eqref{eq:x}
     \State Estimate $\hat{\theta}_{\ell}$ via MUSIC on the signal $\vv{x}$
     \EndFor
     \State Define $\mathcal{A}^{(n+1)}$ as the union of the subareas corresponding to the $N_A -1$ $\{\hat{\theta}_{\ell}\}_{\ell=1}^{N_A}$ that maximize the likelihood
     \State Set $\hat{\theta}^{(n+1)}$ as the one that maximizes the likelihood among the avaialable $\{\hat{\theta}_{\ell}\}_{\ell=1}^{N_A}$
     \State Divide  $\mathcal{A}^{(n+1)}$ into $N_A$ equal subareas $\{\mathcal{A}^{(n+1)}_1,\ldots,\mathcal{A}^{(n+1)}_{N_A}\}$ with $\Delta$ overlap
      \State $n \gets n+1$
     \EndWhile
     \State Calculate $\vv{R}_{\vv{y}s}$
     \State Set $m^{\star}$ as the lag corresponding to the peak in $\vv{R}_{\vv{y}s}$
     \State Set $\tau = m^{\star}/\Delta f$
     \State Fix $\hat{\theta}=\hat{\theta}^{(n)}$ and $\hat{d} = \tau c - d_{\rmG}$
  \end{algorithmic}
\end{algorithm}



\section{Numerical Results}

In this section, we test the localization performance of \name{} in a realistic scenario, in which we assume the \gls{ris} to be a squared structure with $N_x = N_y$ elements. \change{For ease of presentation, we consider the \gls{ap}, the \gls{ris} and the \gls{ue} to be on the horizontal plane $z = 0$.} The simulation parameters are listed in Table~\ref{tab:parameters}, unless otherwise stated. 

\begin{table}[h!]
\caption{Simulation settings}
\label{tab:parameters}
\centering
\resizebox{\linewidth}{!}{%
\begin{tabular}{cc|cc|cc}
\textbf{Parameter} & \textbf{Value} & \textbf{Parameter} & \textbf{Value} & \textbf{Parameter} & \textbf{Value}\\  
\hline
\rowcolor[HTML]{EFEFEF}
$M$             & 4                 & $N_x$, $N_y$           & 4        & $d_{\rmG}$    & 50 m\\
$\psi_{D,x}$    & 225\textdegree    & $\psi_{D,z}$  & 0         & $\psi_A$      & 45\textdegree  \\
\rowcolor[HTML]{EFEFEF}
 $P$            & 20 dBm            & $\sigma$    & -80 dBm   & $\beta$       & 2\\
 $\epsilon$            & $0.5^{\circ}$                 & $N_A$         & 3         & $\delta$      & 0.5 \\ 
 \rowcolor[HTML]{EFEFEF}
 $\Delta f$            & 30.72 MHz            &    $T$   & 10$^3$ & $\Delta$ & 0.2\\

\end{tabular}%
}
\end{table}

\begin{figure}[t!]
    \centering  
    \includegraphics[clip,width=\linewidth]{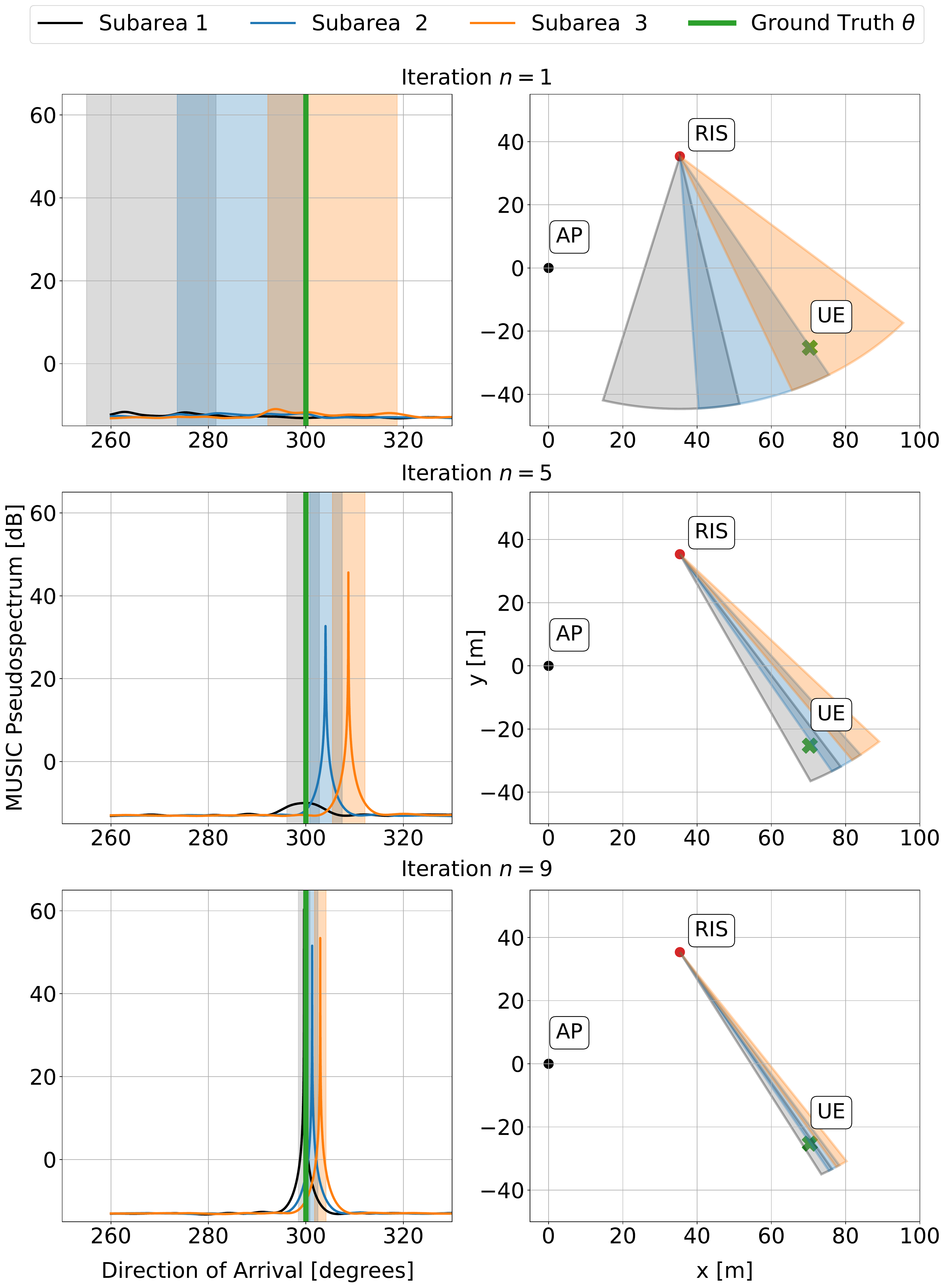}
    \caption{One instance of \name{} \gls{doa} estimation.}
    \label{fig:time_evolution}
\end{figure}

\begin{figure}[t!]
    \centering  
    \includegraphics[clip,width =  \linewidth]{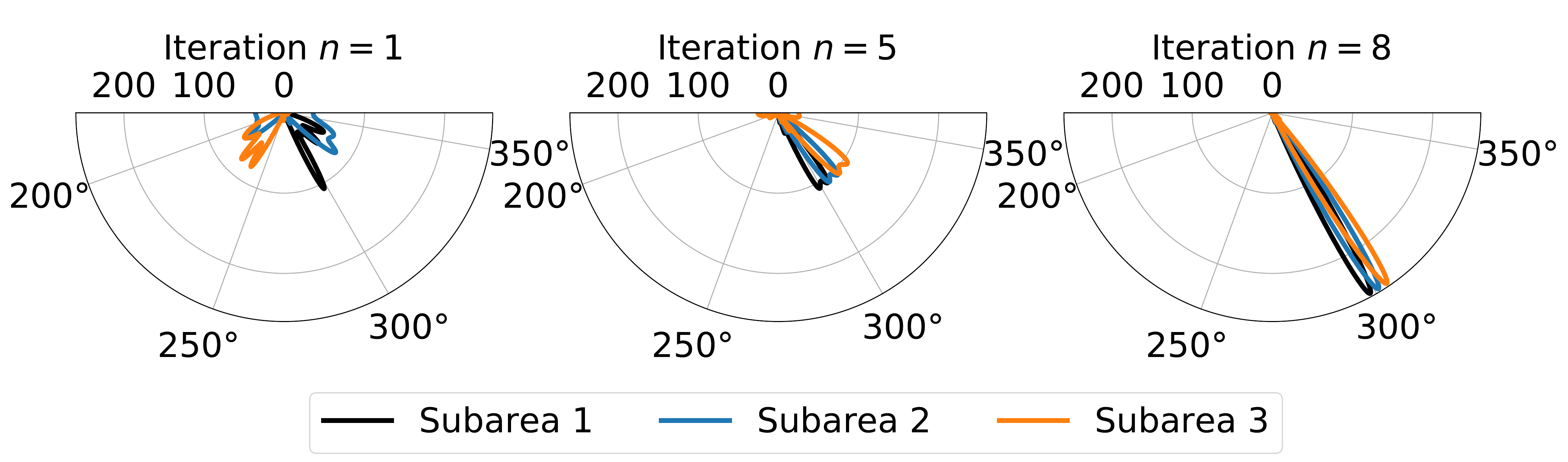}
    \caption{Antenna diagram of the \gls{ris} configuration for different iterations of one instance of the \name{} \gls{doa} estimation.}
    \label{fig:polarplot}
\end{figure}

Fig.~\ref{fig:time_evolution} shows three iterations of an instance of \name{} \gls{doa} estimation. In this example, the \gls{ue} is at coordinates $\vv{p}$ = $[d \cos(\theta), d \sin(\theta), 0]^\tran$, where $\theta \equiv \theta_x$ and $\theta_z = 0$, with $\theta_x = 300^\circ$ and $d = 70$ m. As expected, the narrower the probe areas the sharper the peaks in the respective MUSIC pseudospectrum. In particular, the last iteration returns an estimate $\hat{\theta}$ with a  corresponding absolute \gls{doa} error of less than one tenth of a degree. It is worth pointing out that the pseudospectrum peak is not necessarily associated to a high estimation accuracy due to its dependency on the beamforming optimization at the \gls{ris}, whose antenna diagrams are respectively depicted in Fig.~\ref{fig:polarplot}. Not surprisingly, the beamformers become iteratively more selective, although their overall shape is not regular, especially in the first iterations. This further motivates the choice of leveraging the prior statistical information on the \gls{ue} position rather than the amplitude of the MUSIC pseudospectrum peaks, which may be affected by beamforming artifacts.

\begin{figure}[t!]
    \centering  
    \includegraphics[clip,width = 0.85 \linewidth]{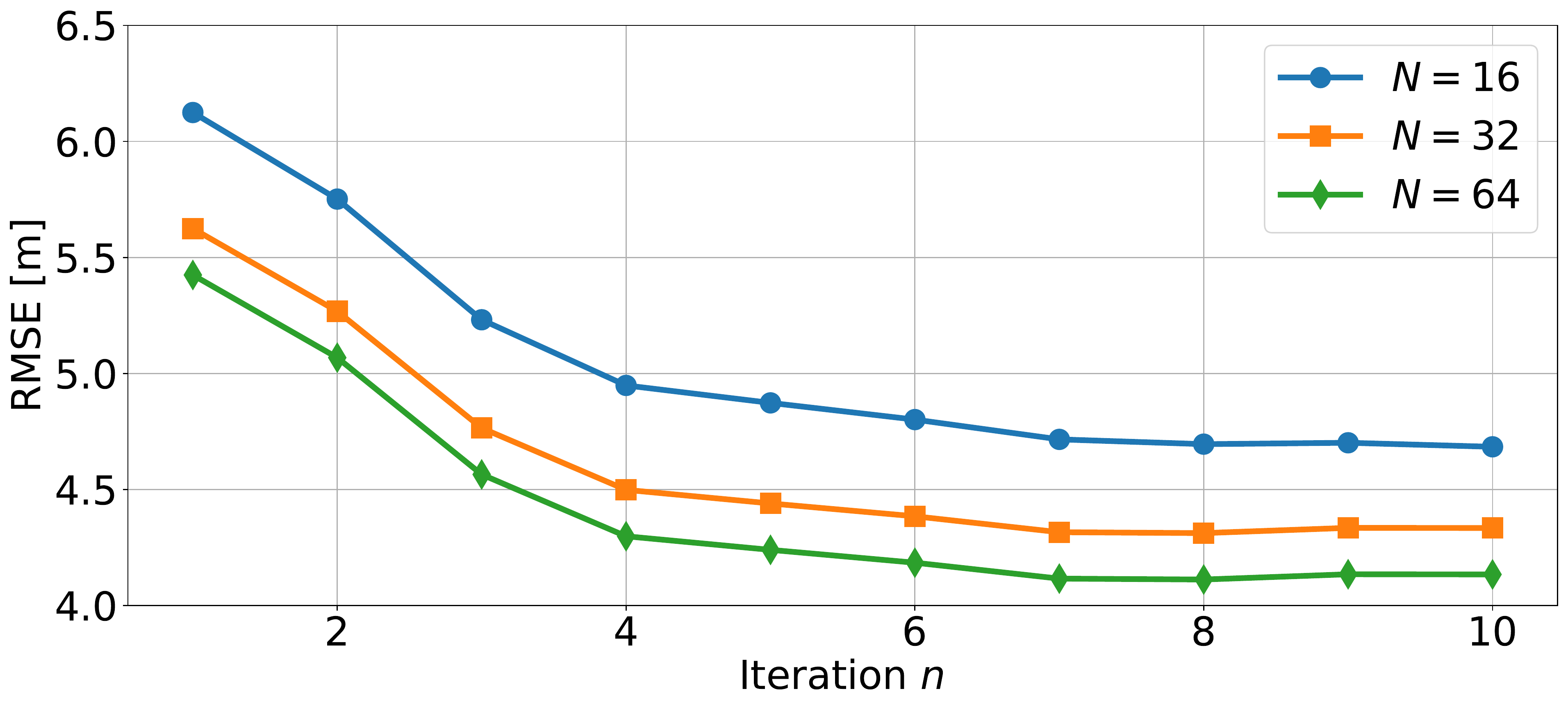}
    \caption{Overall \name{} localization accuracy in terms of RMSE against iterations.}
    \label{fig:error_montecarlo}
\end{figure}

Moreover, we investigate the \name{} overall localization accuracy in terms of \gls{rmse}. To this aim, we assume that the \gls{ue} position is uniformly distributed as $f_{\rmpp} (\vv{p}) = \mathcal{U}[260^\circ,320^\circ]\times\mathcal{U}[20 \, \mathrm{m},80\,  \mathrm{m}]$ and sample such distribution by drawing $T$ points. Note that we average our simulation results over $10^3$ Monte Carlo runs. In Fig.~\ref{fig:error_montecarlo}, we depict the \gls{rmse} performance against different numbers of \gls{ris} elements $N \in \{16,32,64\}$ over the iterations of the \name{} algorithm. Although the \gls{rmse} decreases with the number of iterations for any $N$, the relative gain obtained by increasing $N$ appears to have a diminishing trend, which suggests the existence of a sweet spot in the \gls{ris} configuration for this particular localization problem. 

\section{Conclusions}

In this paper we presented \name{}, a practical localization system leveraging on \glspl{ris},
which estimates \gls{ue} positions by performing statistical beamforming, \gls{doa} and \gls{toa} estimation on a given three-dimensional search space.    
To the best of our knowledge, \name{} is the first two-stage localization solution employing a fully-passive single-\gls{ris} in reflection mode. \name{} takes advantage of prior statistical information on the \gls{ue} position to optimize the \gls{ris} reflection coefficients as to probe a given search space, which is iteratively updated on the basis of the likelihood of the \gls{ue} position.

\section{Acknowledgements}
This work has been supported by EU H2020 RISE-6G project (grant number 101017011).

\bibliographystyle{IEEEtran}
\bibliography{IEEEabrv,bibliography}

\begin{thebibliography}{10}
\providecommand{\url}[1]{#1}
\csname url@samestyle\endcsname
\providecommand{\newblock}{\relax}
\providecommand{\bibinfo}[2]{#2}
\providecommand{\BIBentrySTDinterwordspacing}{\spaceskip=0pt\relax}
\providecommand{\BIBentryALTinterwordstretchfactor}{4}
\providecommand{\BIBentryALTinterwordspacing}{\spaceskip=\fontdimen2\font plus
\BIBentryALTinterwordstretchfactor\fontdimen3\font minus
  \fontdimen4\font\relax}
\providecommand{\BIBforeignlanguage}[2]{{%
\expandafter\ifx\csname l@#1\endcsname\relax
\typeout{** WARNING: IEEEtran.bst: No hyphenation pattern has been}%
\typeout{** loaded for the language `#1'. Using the pattern for}%
\typeout{** the default language instead.}%
\else
\language=\csname l@#1\endcsname
\fi
#2}}
\providecommand{\BIBdecl}{\relax}
\BIBdecl

\bibitem{DiRenzo2020}
M.~{Di Renzo} \emph{et~al.}, ``{Smart Radio Environments Empowered by
  Reconfigurable Intelligent Surfaces: How It Works, State of Research, and The
  Road Ahead},'' \emph{{IEEE} J. Sel. Areas Commun.}, vol.~38, no.~11, pp.
  2450--2525, 2020.

\bibitem{Wu2020}
Q.~{Wu} and R.~{Zhang}, ``{Towards Smart and Reconfigurable Environment:
  Intelligent Reflecting Surface Aided Wireless Network},'' \emph{{IEEE}
  Commun. Mag.}, vol.~58, no.~1, pp. 106--112, 2020.

\bibitem{Mur20}
P.~{Mursia} \emph{et~al.}, ``{RISMA: Reconfigurable Intelligent Surfaces
  Enabling Beamforming for IoT Massive Access},'' \emph{{IEEE} J. Sel. Areas
  Commun.}, vol.~39, no.~4, pp. 1072--1085, 2020\color{black}.

\bibitem{Cal21}
E.~Calvanese~Strinati \emph{et~al.}, ``{Wireless Environment as a Service
  Enabled by Reconfigurable Intelligent Surfaces: The {RISE-6G} Perspective},''
  \emph{Proceedings of EuCNC 6G Summit}, 2021.

\bibitem{Mur21}
P.~Mursia \emph{et~al.}, ``{RISe of Flight: RIS-Empowered UAV Communications
  for Robust and Reliable Air-to-Ground Networks},'' \emph{IEEE Open J. Commun.
  Soc.}, pp. 1--1, 2021\color{black}.

\bibitem{Wymeersch2020}
H.~{Wymeersch} and B.~{Denis}, ``{Beyond 5G Wireless Localization with
  Reconfigurable Intelligent Surfaces},'' in \emph{Proc. {IEEE} Int. Conf.
  Commun. (ICC)}, 2020, pp. 1--6.

\bibitem{Wymeersch2020VTM}
H.~{Wymeersch} \emph{et~al.}, ``{Radio Localization and Mapping With
  Reconfigurable Intelligent Surfaces: Challenges, Opportunities, and Research
  Directions},'' \emph{{IEEE} Veh. Technol. Mag.}, vol.~15, no.~4, pp. 52--61,
  2020.

\bibitem{He2020b}
J.~{He} \emph{et~al.}, ``{Adaptive Beamforming Design for mmWave RIS-Aided
  Joint Localization and Communication},'' in \emph{Proc. {IEEE} Wireless
  Commun. and Netw. Conf. (WCNC)}, 2020, pp. 1--6.

\bibitem{Fascista2020}
\BIBentryALTinterwordspacing
A.~Fascista \emph{et~al.}, ``{RIS-aided Joint Localization and Synchronization
  with a Single-Antenna MmWave Receiver},'' 2020. [Online]. Available:
  \url{https://arxiv.org/abs/2010.14825}
\BIBentrySTDinterwordspacing

\bibitem{Hu2018}
S.~{Hu} \emph{et~al.}, ``{Beyond Massive MIMO: The Potential of Positioning
  With Large Intelligent Surfaces},'' \emph{{IEEE} Trans. Signal Process.},
  vol.~66, no.~7, pp. 1761--1774, 2018.

\bibitem{Ma2021}
T.~{Ma} \emph{et~al.}, ``{Indoor Localization With Reconfigurable Intelligent
  Surface},'' \emph{{IEEE} Commun. Lett.}, vol.~25, no.~1, pp. 161--165, 2021.

\bibitem{Guidi2021}
F.~{Guidi} and D.~{Dardari}, ``{Radio Positioning with EM Processing of the
  Spherical Wavefront},'' \emph{{IEEE} Trans. Wireless Commun.}, pp. 1--1,
  2021.

\bibitem{elzanaty2020}
\BIBentryALTinterwordspacing
A.~Elzanaty \emph{et~al.}, ``{Reconfigurable Intelligent Surfaces for
  Localization: Position and Orientation Error Bounds},'' 2020. [Online].
  Available: \url{https://arxiv.org/abs/2009.02818}
\BIBentrySTDinterwordspacing

\bibitem{He2020}
J.~{He} \emph{et~al.}, ``{Large Intelligent Surface for Positioning in
  Millimeter Wave MIMO Systems},'' in \emph{IEEE Veh. Tech. Conf. (VTC)}, 2020,
  pp. 1--5.

\bibitem{Zhang2021}
H.~{Zhang} \emph{et~al.}, ``{Towards Ubiquitous Positioning by Leveraging
  Reconfigurable Intelligent Surface},'' \emph{{IEEE} Commun. Lett.}, vol.~25,
  no.~1, pp. 284--288, 2021.

\bibitem{nguyen2020}
\BIBentryALTinterwordspacing
C.~L. Nguyen \emph{et~al.}, ``{Reconfigurable Intelligent Surfaces and Machine
  Learning for Wireless Fingerprinting Localization},'' 2020\color{black}.
  [Online]. Available: \url{https://arxiv.org/abs/2010.03251}
\BIBentrySTDinterwordspacing

\bibitem{Abeywickrama2020}
S.~Abeywickrama \emph{et~al.}, ``{Intelligent Reflecting Surface: Practical
  Phase Shift Model and Beamforming Optimization},'' \emph{{IEEE} Trans.
  Commun.}, vol.~68, no.~9, pp. 5849--5863, 2020.

\end{thebibliography}

\end{document}